# The Impact of COVID-19 Pandemic on Ridesourcing Services Differed Between Small Towns and Large Cities

**Nael Alsaleh[1] and Bilal Farooq[1,*]**

[1]Laboratory of Innovations in Transportation (LiTrans), Department of Civil Engineering, Ryerson University, Toronto, ON M5B 2K3, Canada
[*]bilal.farooq@ryerson.ca

## ABSTRACT

The COVID-19 pandemic has significantly influenced all modes of transportation. However, it is still unclear how the pandemic affected the demand for ridesourcing services and whether these effects varied between small towns and large cities. We analyzed over 220 million ride requests in the City of Chicago (population: 2.7 million), Illinois, and 52 thousand in the Town of Innisfil (population: 37 thousand), Ontario, to investigate the impact of the COVID-19 pandemic on the ridesourcing demand in the two locations. Overall, the pandemic resulted in fewer trips in areas with higher proportions of seniors and more trips to parks and green spaces. Ridesourcing demand was adversely affected by the stringency index and COVID-19-related variables, and positively affected by vaccination rates. However, compared to Innisfil, ridesourcing services in Chicago experienced higher reductions in demand, were more affected by the number of hospitalizations and deaths, were less impacted by vaccination rates, and had lower recovery rates.

## Introduction

Since 2011, smartphone App-based ridesourcing services have emerged as a new travel mode in many cities worldwide[1–5]. Over the past decade since its inception, the service has seen a dramatic increase in popularity and market share due to features like the flexibility in the pickup and dropoff locations and times, safety, wide coverage areas, and reliability[5–8]. However, in March 2020, the 2019 novel coronavirus (COVID-19) began to spread rapidly across the globe, prompting the World Health Organization (WHO) to declare it a global pandemic[9–11]. Consequently, governments have introduced various Non-Pharmaceutical Interventions (NPIs) to curb the spread of the virus including lockdowns, stay-at-home orders, work-from-home policies, closing schools, universities, and non-essential workplaces, social distancing, wearing masks, and movement restrictions[12–15]. Additionally, by the beginning of 2021, different COVID-19 vaccines have been approved for emergency use by the WHO and governments have started to immunize their population to reduce the risk of getting and transmitting the virus. Nevertheless, after more than two years since the first case was reported, COVID-19 is still an ongoing pandemic mainly due to the vaccine-hesitancy and inequality as well as the emergence of new variants[16,17]. This raises an important question, how did the COVID-19 pandemic impact the transportation demand in general and the demand for ridesourcing services in particular?

Previous literature revealed that the pandemic has significantly influenced all modes of transportation as well as changed individual's travel behaviour and preferences[18–22]. During the COVID-19 first and second waves, public transit, taxi, and ridesourcing services have experienced substantial reductions in the ridership[20,23,24]. However, public transit users were more likely to change their travel mode compared with the users of other modes, due to their concerns regarding the safety of the service and fears of getting infected[19,20,25–27]. Moreover, user's low confidence in public transit is expected to slow down the recovery of the system[28]. On the contrary, the use of private vehicles, motorcycles, and active means of transportation have become more common[20,25].

The reduction of transportation demand during the early stages of the pandemic was mainly due to the COVID-19 related variables, e.g., new hospitalized cases, NPIs, as well as individual's attitudes and perceptions towards the pandemic. A recent study in China explored the impact of the pandemic on the behaviour of ridesourcing drivers using the actual data from September 2019 to August 2020. The study indicated that the daily number of new COVID-19 cases significantly affected the number of trips drivers made at the beginning of the pandemic. However, the effect of newly recorded cases decreased during the reopening phase[24]. Another recent study in the same country showed that the demand for both ridesourcing and taxi services was significantly affected by the number of COVID-19 cases, COVID-19 related policy measures, mean transportation cost, and the operational status of mass transit[20]. In the United States, social distancing measures and stay-at-home orders implemented in March 2020 resulted in nearly 30% reduction of personal trips[14]. The impact of the COVID-19 related interventions at the beginning of the pandemic was not limited to the motorized modes, but they have also significantly reduced the daily walking behaviour of people in many places. However, in the subsequent months, when the weather became warmer and some commercial activities resumed, recreational walking behaviour has increased markedly[29].

On the other hand, it was documented that the risk attitude of individuals was more influential on mobility than the actual mortality and hospitalization rates of COVID-19, especially during the early stages of the pandemic[30]. Using data obtained from a web-based survey in Greater Toronto Area (GTA) in July 2020, researchers showed that the perception of risk and safety has affected individual's decision of using ridesourcing services[23]. In another recent study, the authors examined the travel behaviour changes at the beginning of the pandemic in Java Island, Indonesia, using a web-based survey data. The results revealed that individuals who took preventative measures for the COVID-19 when leaving home were less likely to engage in outdoor activities and more likely to reduce the frequency of using ridesourcing services[31].

To summarize, various NPIs have been adopted to reduce individual's mobility which, in turn, played a major role in slowing the transmission of the virus[32, 33]. In a recent study, researchers used mobility data along with web search queries to identify the potential locations of COVID-19 outbreak hotspots[34]. Other studies simulated the impact of individual's mobility and vaccination rates on the evolution of the pandemic[35] or provided policy options for the recovery of public transit and shared mobility[36, 37]. Yet, it is still unclear whether the effects of the pandemic on ridesourcing services differed between small towns and large cities. Moreover, the implemented NPIs and their strictness varied across countries and over time based on the epidemiological situation, the availability of the vaccines, vaccination rates, the existence of new variants, and the economic situation. Therefore, there is a strong need to examine the effects of the severity of NPIs and vaccination rates on the reduction in ridesourcing demand.

In the current study, COVID-19 related variables and NPIs, land-use data, socioeconomic and demographic characteristics, weather data, and the actual ridesourcing data for the City of Chicago, Illinois, and the Town of Innisfil, Ontario, from November 2018 to August 2021 are used to analyze and compare (a) the impacts of the COVID-19 pandemic on the spatio-temporal patterns of the demand, (b) the effects of the severity of the NPIs, COVID-19 related variables, and vaccination rates on the percent reduction in the daily demand, as well as (c) the main factors affecting the direct demand (origin-destination-pair) in the post-pandemic era. This work can be of use to policymakers as well as service providers to understand the future impact of the pandemic on ridesourcing services in terms of the recovery of the service, preparedness for additional waves of COVID-19 as well as other pandemics.

## Methods

### Data collection

The Town of Innisfil is located on the western shore of Lake Simcoe in Simcoe County, Ontario, with a population density of 139 people per square kilometer. According to the 2016 census data, the town has a population of 36,566 people, around 67% of whom are in the working age group (15 to 64 years old). The median household income (after-tax) in Innisfil is $57,846, which is higher than the national average of $48,895[38]. On the other hand, the City of Chicago is located on the southwestern tip of Lake Michigan in Illinois and is considered among the largest cities in the United States, with a population density of 4,618 persons per square kilometer. The City of Chicago has a population of more than 2.7 million and almost 70% of the population belongs to the working age group. The median household income (after-taxes) in Chicago is $62,613, which is lower than the national average of $70,690[39].

The town of Innisfil uses ridesourcing services as its primary transit service, since it partnered with the ridesourcing company, Uber Technologies Inc., to provide residents with subsidized trips instead of operating a fixed-route transit system[40]. Therefore, the service has been used for a wide variety of trip purposes including work, school, shopping, social, recreation, and medical appointments[41]. On the other hand, ridesourcing services are one of the main modes of travel in Chicago that operate independently of the public transportation service and are used mainly for non-work purposes[42]. We hypothesize that the differences in use purposes, operating policies, land area, and population density resulted in the COVID-19 pandemic having different effects on ridesourcing services between the two locations.

In this study, data from multiple sources were utilized to examine this hypothesis. Weather data were collected from AccuWeather Inc.[43] for both case studies. The collected dataset contained information on the daily average temperature, precipitation, and snowfall for the period of March 1, 2020 to July 31, 2021. Socioeconomic and demographic factors were obtained from the 2016 Statistics Canada and the 2017 American Community Survey for Innisfil and Chicago, respectively. Several characteristics were collected for both case studies at the census tract level (CT), including population density, median income, gender (percentage of males), percentage of elderly people, education, marital status, and mode of commuting. Land-use data were obtained from the Chicago Metropolitan Agency for Planning and the Town of Innisfil. The COVID-19 related variables were acquired from the Simcoe Muskoka District Health Unit and Chicago COVID Dashboard for the period of January 1, 2020 to July 31, 2021. The main COVID-19 related variables considered in this study were the daily confirmed cases, deaths, hospitalizations, test positivity rate (percentage of tests performed that were positive for COVID-19), as well as the cumulative percentage of residents who received the first and second vaccine doses. As for the severity of the COVID-19 NPIs, we used Stringency Index data from the Oxford COVID-19 Government Response Tracker (OxCGRT). The index represents the strictness of a government's policy measures taken over time to restrict individual's movements and behaviour, and is calculated using nine indicators, including: (a) school and university closure, (b) workplace closure, (c) public event cancellations and restrictions, (d) restrictions on private gatherings, (e) public transport closures, (f) stay-at-home orders, (g) restrictions on intra-provincial travel, (h) restrictions on international travel, and (i) public information campaigns[44]. This dataset was collected for both Innisfil and Chicago from January 1, 2020 to July 31, 2021.

The actual ridesourcing trip data for Chicago were acquired from Chicago Data Portal[45] form November 1, 2018 to July 31, 2021, including 220,527,209 observations (ride requests). Each observation contained, amongst others, information about trip request date, pickup and dropoff times and CT, as well as trip distance, duration, and fare. On the other hand, the actual ridesourcing data for the second case study was provided through Uber Technologies, Inc. and the Town of Innisfil. The data obtained from Uber were for the pre-pandemic period and included information about the average hourly demand for a typical weekday and weekend as well as the average daily demand per month from May 2017 till February 2020. However, for consistency, only the data from November 2018 till February 2020 were used in this study. Moreover, the average hourly demand data were provided for 5 time periods, as follows:

- Morning, from 6:00 AM to 10:00 AM

- Midday, from 10:00 AM to 3:00 PM

- Late afternoon, from 3:00 PM to 7:00 PM

- Evening, from 7:00 PM to 10:00 PM

- Night, from 10:00 PM to 6:00 AM

The Town of Innisfil provided us with the operational data for Uber service over the period of September 1, 2020 to July 31, 2021, including 52,126 trip requests. The operational dataset contained detailed information on trip request time, pickup and dropoff times and location, as well as trip distance, duration, and fare. Note that the dataset does not cover the first wave of the COVID-19 outbreak; however, it can be useful to capture the effects of both the second and third waves, which had higher impacts on the healthcare system, as well as the vaccination rates on the ridesourcing service in Innisfil.

**Data analysis and modelling**
Fig.1 demonstrates the methodological framework used in this study. As Figure 1 depicts, the current study involved three main parts. In the first part, we used the actual ridesourcing data to compare the impacts of the COVID-19 pandemic on the ridesourcing services in the Town of Innisfil and the City of Chicago in terms of the hourly, daily, and monthly distributions of the demand as well as the origin-destination (OD) flow patterns. In the second part, we explored the impact of the COVID-19 stringency index on the daily and hourly demand values. Moreover, the actual ridesourcing data as well as the COVID-19 related variables and stringency index were used to model the percent reduction in the daily demand for the ridesourcing services in Innisfil and Chicago. In both models, the dependent variable represented the percent reduction in the daily demand values at the town/city level, which were computed based on the average weekday and weekend demand values for the pre-pandemic period (from November 1, 2018 to February 28, 2020). On the other hand, the explanatory variables used in each model included the COVID-19 related variables and stringency index as well as the key trip characteristics, such as: the day-of-week, month-of-year, and weekend dummy variable (to distinguish between weekdays and weekends). In the third part, we modelled the daily direct demand for ridesourcing services in each case study during the pandemic at CT level using COVID-19 related variables and stringency index, land-use data, socioeconomic and demographic characteristics, weather data, and the actual ridesourcing data. In both models, the dependent variable was formed by aggregating trips data, from March 1, 2020 to July 31, 2021 for the City of Chicago and from September 1, 2020 to July 31, 2021 for the Town of Innisfil, at the OD-pair level on each day.

The percent reduction in the daily demand and the direct demand models were developed using the Random Forest (RF) regression algorithm. In all models, data were randomly divided into training and test sets with 4:1 ratio. Moreover, Bayesian optimization approach was used to find the optimal architecture of each model with the lowest mean-square error for cross-validation (Supplementary Table 1). The optimal architecture of each model was applied on the test set. Four performance metrics were used to evaluate the predictive accuracy of the models including Mean Absolute Error (MAE), Mean Squared Error (MSE), Root Mean Squared Error (RMSE), and Coefficient of Determination ($R^2$). Furthermore, we used a post-hoc model interpretability method, SHapley Additive exPlanations (SHAP)[46], to examine the impact of the explanatory variables and their relative importance[47,48].

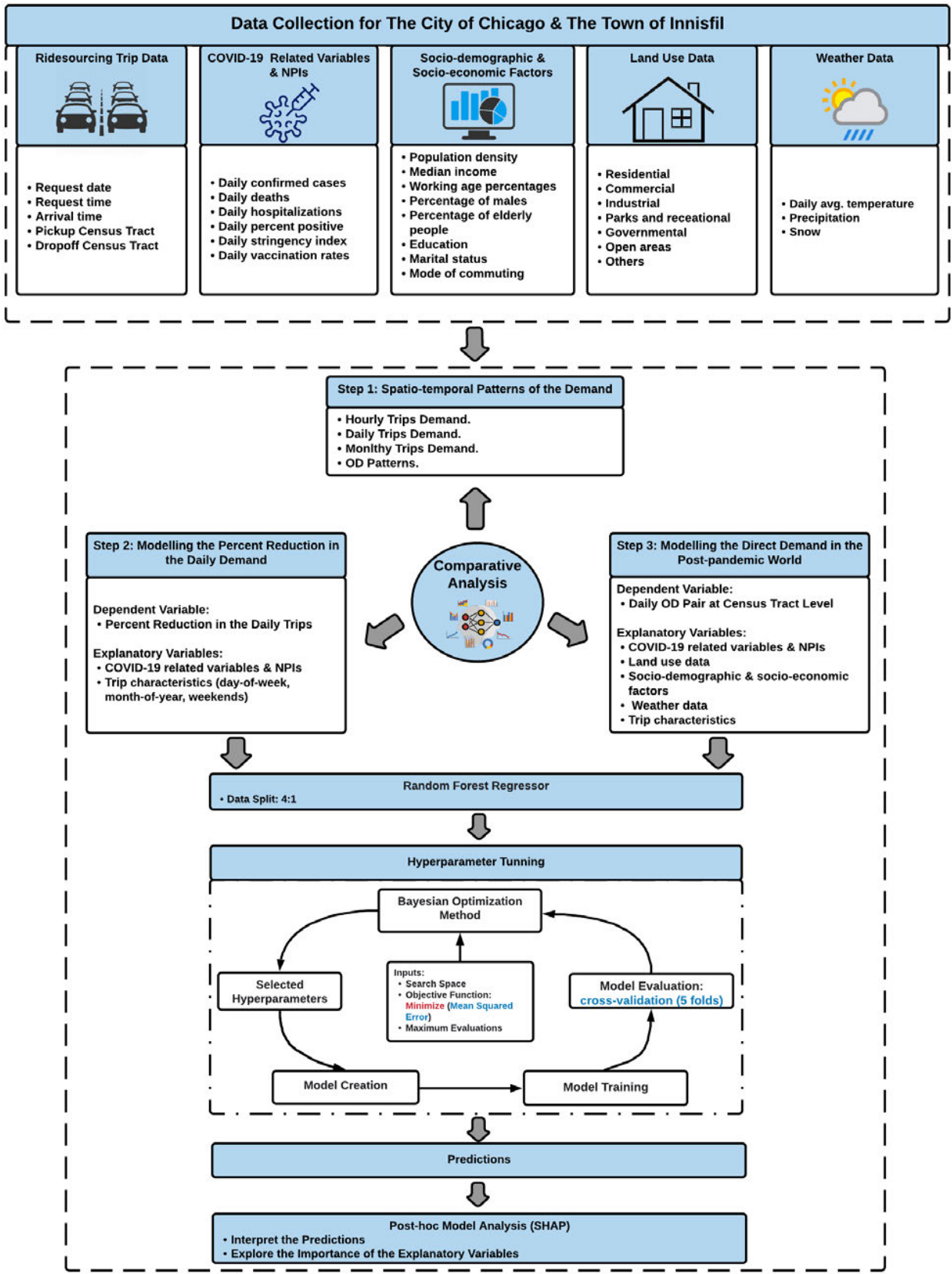

**Figure 1.** Research Framework

## Results

### Spatio-temporal demand patterns

Fig. 2 presents the hourly, daily, and monthly distributions of the demand for ridesourcing services in Innisfil and Chicago pre and during the COVID-19 pandemic. Overall, it is observed that the demand in Chicago is much higher than in Innisfil. This is most likely due to the considerable difference in the population density and land area as well as in the attractions and recreational activities available in both locations. The average daily demand in Chicago was about 203,365 trips in the pre-pandemic period and 45,127 trips during the pandemic, while in Innisfil, the average values were about 216 and 153 trips per day for the pre and during pandemic periods, respectively. Before the pandemic, the demand was highest during the late afternoon period in Innisfil and during the late afternoon and evening periods in Chicago on both weekdays and weekends. Among the days of the week, the demand was the highest during the weekdays in Innisfil and during the weekends in Chicago. These patterns remained the same during the pandemic. However, the pandemic had a strong impact on the demand levels, and these effects varied significantly between the two locations.

On average, the hourly and daily demand values in Innisfil were reduced by 30% during the pandemic, whereas the hourly and daily demand values in Chicago were 80% less compared with the pre-pandemic levels. This might be due to the difference in the use purposes of the ridesourcing services between the two locations. The highest reduction in the demand in Innisfil on both weekdays and weekends was during the night period. Moreover, among the days of the week, weekends experienced higher reduction level in the demand. In Chicago, however, we observed a consistent and sharp decline in demand across all times of the day and across all days of the week. Even though the pandemic affected all trips, these findings indicate that non-work related trips were most impacted by the closure of non-essential businesses. It can also be noticed that the reduction levels in the monthly demand varied overtime. Both locations encountered a repeated pattern of a sharp decrease in the demand followed by a gradual recovery. In Innisfil, the highest reduction in the monthly demand was observed in January 2021 during the second wave of COVID-19 at 52.9% and the lowest in July 2021 following the end of the third wave at 8.5%. On the other hand, Chicago experienced the highest decrease in the monthly demand during the first wave in April 2020, at 97.5%, and the lowest in March 2020 at the beginning of the pandemic, at 58.6%.

Fig. 3 displays the origin-destination flow patterns for the ridesourcing services in Innisfil and Chicago during the pandemic. In Innisfil, most of the OD pairs with high demand were found within the central area as well as between the central area and the neighbouring CTs containing the train stations. These patterns are not surprising, since the central area has the highest population density and contains several shopping malls, office buildings, and large grocery stores. On the other hand, the OD pairs associated with the train stations are likely to be the morning and afternoon commutes. In Chicago, the high-demand OD pairs were within the downtown area and between the downtown area and O'Hare International Airport. Chicago's downtown area is characterized by high population density, median income, and employment levels. The area is also home to large shopping malls, parks, hotels, as well as recreation and commercial areas. These factors have a positive effect on the ridesourcing demand[49,50]. O'Hare, one of the busiest and largest airports in the world, is expected to be one of the most popular destinations for ridesourcing. Finally, the pre-pandemic ridesourcing trips showed the same patterns in both locations (Supplementary Fig. 1).

### Factors influencing the percent reduction in the daily demand

Here, we investigate the effects of the COVID-19 related variables, government stringency index, and the key trip characteristics on the daily demand for the ridesourcing services. Fig. 4a illustrates the relationship between the stringency index, vaccination rates, COVID-19 cases, and the daily trip demand in Innisfil and Chicago. In general, the increase in the stringency index values was associated with a decline in the daily trip demand as well as the COVID-19 cases reported 2 weeks later. Ridesourcing services saw a significant decrease in demand immediately after each jump in the stringency index, followed by a gradual recovery. However, the extent of the stringency index effect varied among the two locations, and it had a noticeably higher impact on the ridesourcing demand in Chicago. Moreover, ridesourcing demand in Chicago showed a lower recovery rate, when compared to that in Innisfil. We argue that these trends are explainable by the differences in the use purposes and the operating policy between the two locations. High stringency index values have primarily restricted non-mandatory trips which represent a high proportion of the total trips served by the ridesourcing services in Chicago. Furthermore, it is noticed that the increase in the daily vaccination rates, first and second vaccine doses, was correlated with an increase in the ridesourcing trip demand and a decrease in the COVID-19 cases in both locations.

The impact of the stringency levels (less than 25, 25-50, 50-75, and more than 75) on the weekdays and weekends hourly demand values are shown in Fig. 4b. We found that the hourly demand values continuously reduced with the increase in the stringency level. The extent of the stringency index effect was homogeneous across the day periods during weekdays and higher during the night period on weekends. Moreover, the reduction amount in the hourly demand was consistent with the transition from one stringency level to the next in Innisfil, whereas the second stringency level had the highest impact on the hourly demand values in Chicago.

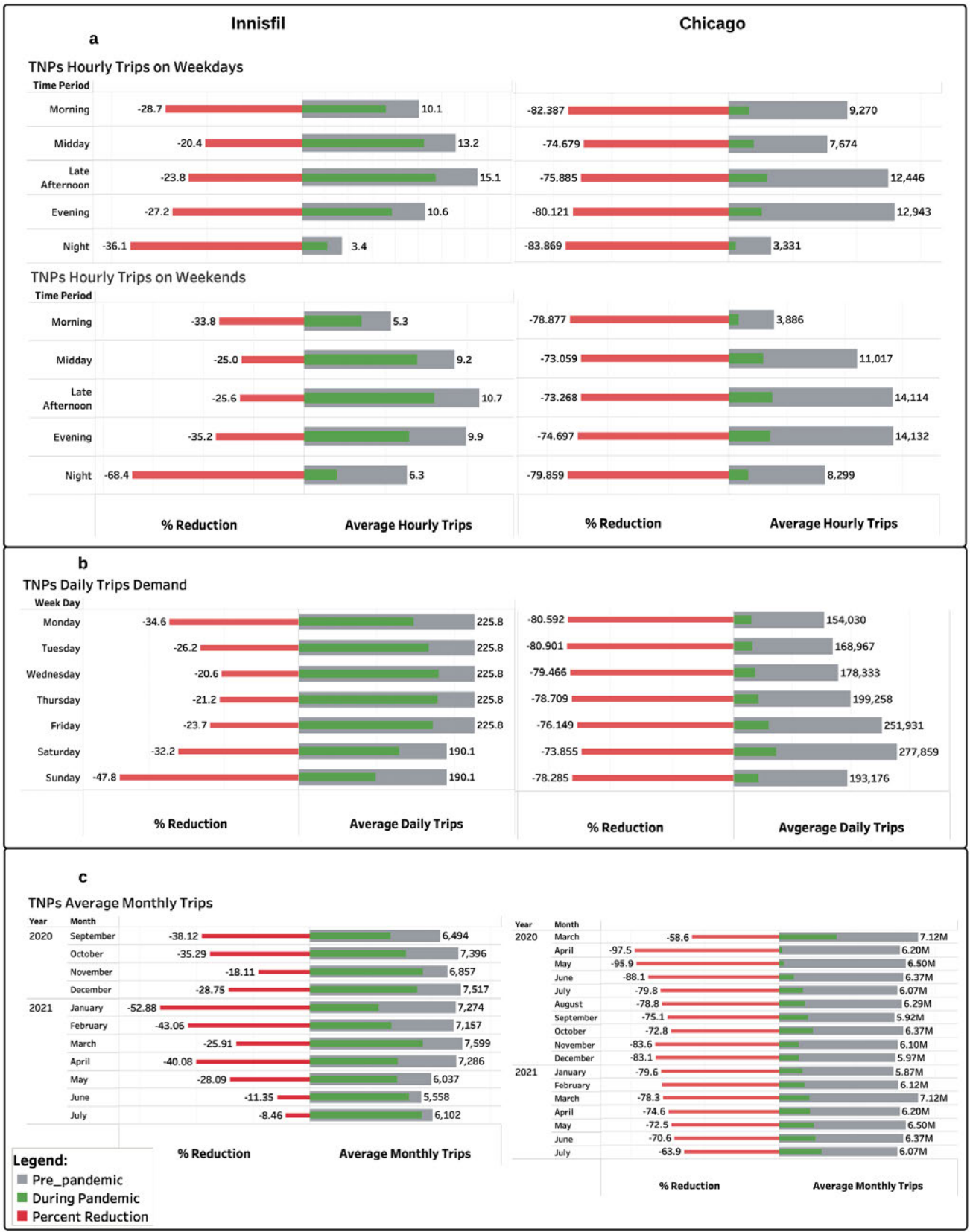

**Figure 2.** COVID-19 pandemic impacts on the temporal distribution of the demand for the ridesourcing services in the Town of Innisfil and the City of Chicago. **(a)** Hourly distribution of the demand pre and during the pandemic. **(b)** Daily distribution of demand pre and during the pandemic. **(c)** Monthly distribution of demand pre and during the pandemic.

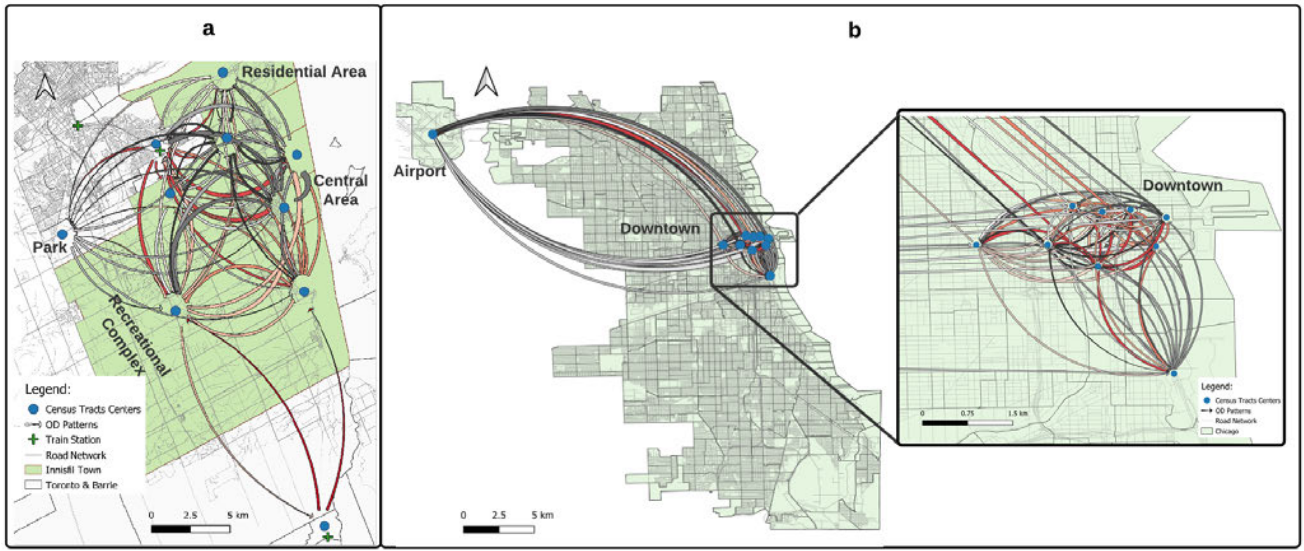

**Figure 3.** Origin-destination (OD) flow patterns for ridesourcing services during the pandemic in **(a)** Innisfil and **(b)** Chicago. In Innisfil, 52,126 ridesourcing trips were made from September 2020 till August 2021 covering 20 CTs, while more than 21 million trips were made in Chicago during the pandemic distributed over 803 CTs. Since it is hard to develop meaningful flow patterns for all OD pairs, we only display the flow patterns for the 10 most frequently used CTs in both locations. The OD flows shown in this figure represent 90% and 10% of the total trips in Innisfil and Chicago, respectively. The OD line color represents the origin of trips and the width reflects the flow strength. The wider the line is, the more trips the OD pair has.

Furthermore, the percent reduction in daily demand was modelled to get a better understanding of how the trip characteristics as well as the COVID-19 related variables and stringency index have affected the daily demand in Innisfil and Chicago (Supplementary Table 2). Fig. 5 presents the SHAP summary plots for the percent reduction in the daily demand models. The results revealed that the stringency index, COVID-19 deaths, hospitalizations, and test positivity rate variables had a negative effect on the demand reduction levels in both locations. Daily demand values experienced higher reduction levels as the stringency index, COVID-19 deaths, hospitalizations, and test positivity rate increased. Regarding the effects of the trip characteristics variables, weekend variable was found to have a negative impact, while both the day-of-week and month-of-year variables had a positive effect on the ridesourcing demand. These findings indicate that the ridesourcing demand saw higher reduction levels on weekends as well as at the beginning of the week and the year. Vaccination rate variables, on the other hand, had a positive impact on the daily demand values. Daily demand values increased with the increase in the number of partially and fully vaccinated individuals. This suggests the importance of the vaccination for the recovery of the ridesourcing services. As for the number of COVID-19 cases, it was found to have a negative impact on the demand reduction levels in Innisfil and a positive effect in Chicago. This is due to the fact that the demand for ridesourcing in Chicago reached its lowest point at the beginning of the pandemic, when both the number of COVID-19 cases and tests were low. Later on when the city conducted more tests and therefore recorded more cases, the demand increased slightly (see Fig. 4).

In terms of the relative importance of variables, the results indicated that the stringency index values had the highest impact on the daily reduction levels in Innisfil, followed by the trip characteristics, vaccination rates, and COVID-19-related statistics, respectively. In Chicago, the percent reduction in ridesourcing demand was most influenced by the trip characteristics, followed by the stringency index, COVID-19 variables, and vaccination rates, respectively. The relative importance of the vaccination rates between the two locations further explains the slower recovery rate observed in Chicago for the ridesourcing services. Additionally, this suggests that the recovery of the service could be linked to factors other than vaccination coverage, for example, the economic recovery, the efficacy of the vaccine against the new variants, and individual attitudes and perceptions towards the virus. However, further analysis is needed to understand the factors affecting the recovery of the ridesourcing services. It can also be noticed that the test positivity rate variable was the most important among the COVID-19 related variables in Innisfil, while the daily number of hospitalizations and deaths were the least important. In contrast, the opposite pattern can be observed in Chicago. This can be explained by the low number of people hospitalized or who died from the COVID-19 in Innisfil as compared to Chicago.

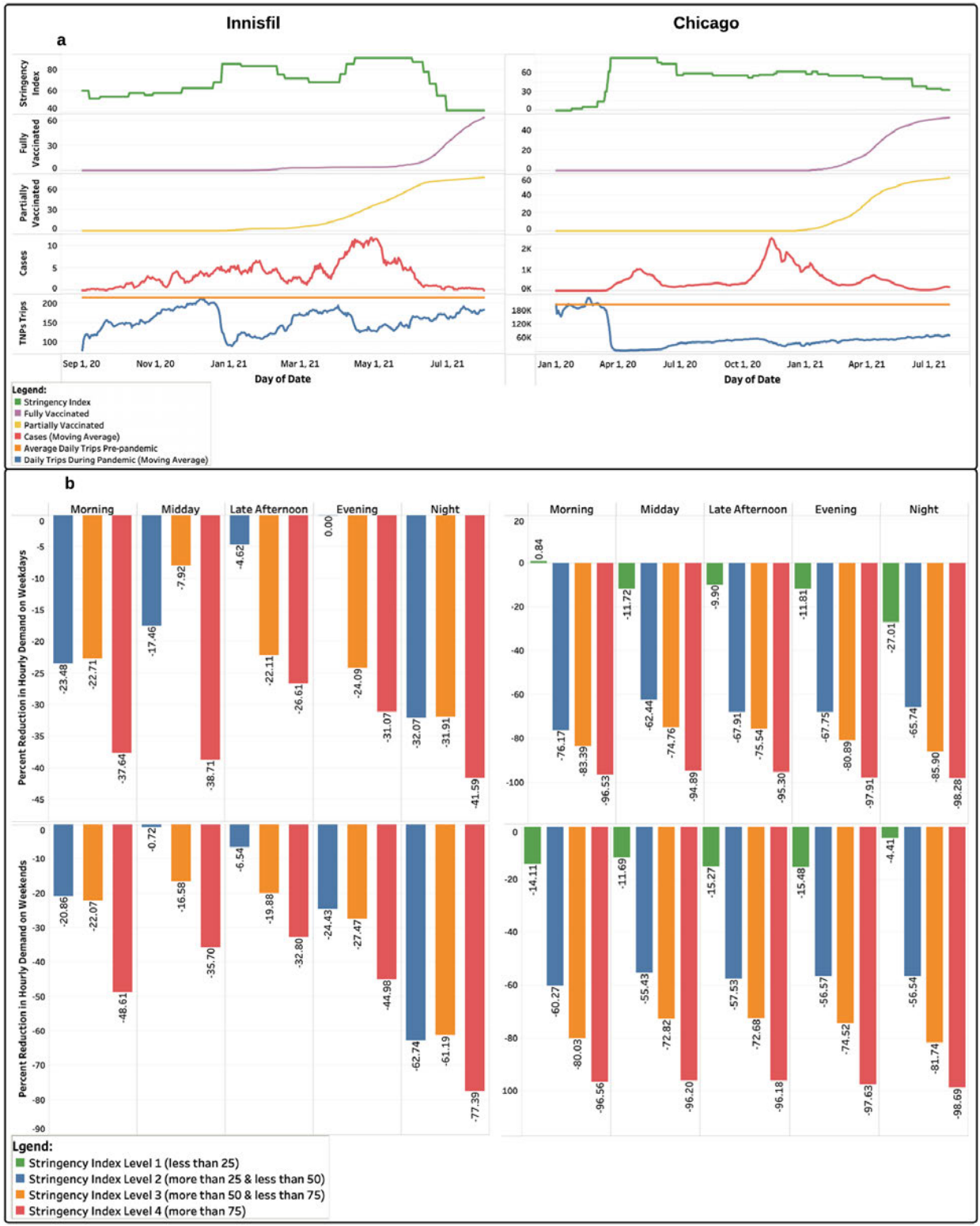

**Figure 4.** (a) The relationship between the stringency index, vaccination rates, COVID-19 cases, and the daily trip demand in Innisfil and Chicago. (b) Effects of stringency index on weekdays and weekends hourly demand values in Innisfil and Chicago. To explore the impact of the stringency index on the hourly demand, we categorized its values into five levels: pre-pandemic (stringency index =0), less than 25, 25-50, 50-75, and more than 75.

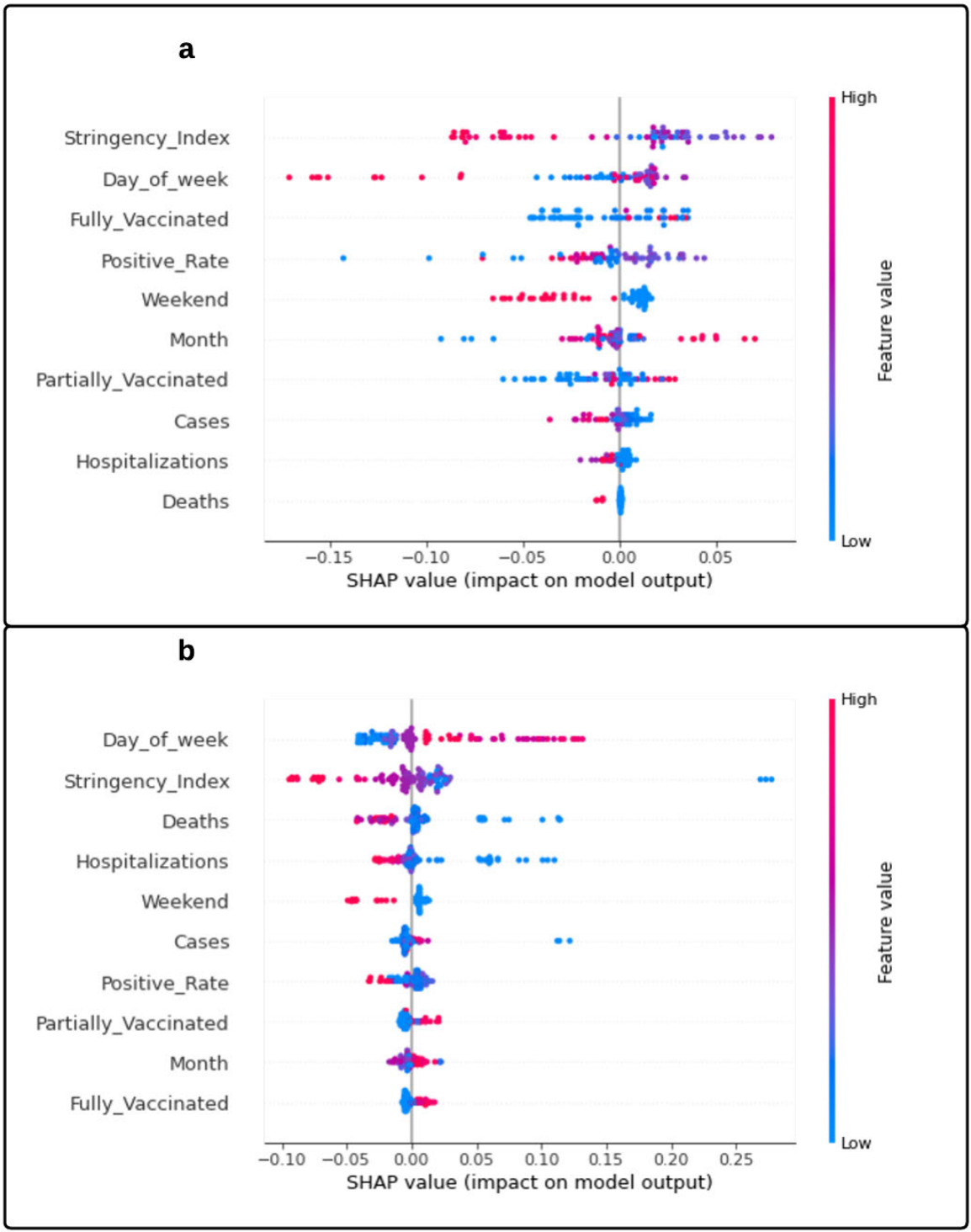

**Figure 5.** SHAP summary plot for the percent reduction in the daily demand models. **(a)** Innisfil model. **(b)** Chicago model. Each point in the summary plot represents a Shapley value for a feature and an instance. The feature importance determines its position on the y-axis, the Shapley value determines its position on the x-axis, and the value determines its colour.

### Factors affecting the daily direct demand for ridesourcing services during the pandemic

Here, we explore the impact of the COVID-19 related variables and stringency index, land-use data, socioeconomic and demographic characteristics, as well as the weather data on the direct demand (OD-pair trips) for the ridesourcing services in Innisfil and Chicago during the pandemic.

### Socioeconomic, demographic, and land-use variables

The results revealed that population density, median income, working-age population, percentage of men, and percentage of married people at the trip origin and destination positively impacted ridesourcing demand in both locations (Figs. 6 and 7). Besides, smaller households, higher levels of education, and lower percentages of workers commuting by personal vehicles at the trip origin and destination were associated with higher demand. These findings are consistent with those of previous studies for pre-pandemic conditions[50,51]. However, household size and education factors showed an opposite pattern in Innisfil. Larger households and lower levels of education were associated with higher demand. This variation is mainly due to the subsidized trips offered by the Town Innisfil, which made the service available for a broader segment of the population. Moreover, the OD-pair demand was negatively influenced by the percentage of elderly people at the trip origin and destination in both locations. This finding is consistent with the previous literature indicating that areas with higher percentage of seniors experienced greater mobility reductions during the pandemic[30]. The main reason behind this is that seniors are more likely to get severe symptoms, be hospitalized, or die from the COVID-19[52].

Furthermore, Innisfil's ridesourcing demand was higher between residential and industrial neighbourhoods, most likely due to work-related trips. In Chicago, the demand for ridesourcing was higher between neighborhoods with governmental and commercial land-use types, probably due to the trips made to and from O'Hare International Airport. Interestingly, the results revealed that large parks in both locations had a high demand for ridesourcing. This is in line with recent findings indicating that the usage of parks and green spaces increased significantly during the pandemic[53]. It is believed that this pattern is related to the fact that during the pandemic, parks were the preferred destination for residents for mental and physical health reasons.

### COVID-19 related variables, vaccination rates, and stringency index

As expected, the results indicated that the COVID-19 related variables and the stringency index adversely affected the OD-pair trips in both Innisfil and Chicago. The frequency of the ridesourcing trips decreased with the increase in the COVID-19 cases, hospitalization, deaths, test positivity rate, and the stringency index values. In contrast, a positive correlation was found between fully and partially vaccinated individuals and ridesourcing demand. These findings point to the importance of incorporating pandemic-related characteristics into travel behaviour studies.

### Trip and weather variables

Ridesourcing demand in both locations was negatively associated with the weekend variable and positively associated with the month-of-year variable. Therefore, OD-pair trips were lower on weekends and at the beginning of the year. Taking into account the demand for ridesourcing services was higher on weekends before the pandemic[50], it may be argued that the pandemic has changed the travel behaviour of individuals. Moreover, the impact of the day-of-week variable varied between the two locations, with a negative impact in Innisfil and a positive effect in Chicago. As revealed in this finding, Innisfil's ridesourcing service appears to be used regularly at the beginning of the week for work-related purposes.

On the other hand, the results showed that the ridesourcing demand increased as temperature decreased. This is in line with previous literature and might be the result of individuals switching from non-motorized modes to ridesourcing in cold weather[50]. Although Innisfil and Chicago have similar weather conditions, the demand in Innisfil decreased under rain and increased under snow, whereas the opposite was true in Chicago. The reason for this might be that in rainy and warm weather, individuals in Innisfil are likely to walk or cycle to nearby destinations, while in snowy and icy weather conditions, they switch to ridesourcing services to avoid slipping or being outside in cold. The increase in demand under rain in Chicago might be related to individuals engaging in more activities in warm weather, and the decrease in snowy conditions might be attributed to event cancellations, lower supply, and higher prices.

### Variable importance

Overall, the socioeconomic, demographic, and land-use variables had the greatest impact on the OD-pair trips in both locations, followed by the trip characteristics and the stringency index, respectively. These results are consistent with the pre-pandemic conditions, where the demand was most affected by the socioeconomic and demographic characteristics[49]. Innisfil's direct demand was least affected by weather factors and COVID-19 characteristics. Among the socioeconomic and demographic characteristics, the average household size at the trip origin and the population density at the trip destination were the most influential factors. Moreover, the test positivity rate had the highest importance among the COVID-19 related variables. In Chicago, vaccination rates and weather variables had the least influence on direct demand. The average household size at the trip origin as well as the education level at the trip destination were the most important socioeconomic and demographic characteristics. Finally, the daily number of cases was the most influential among the COVID-19 related variables. These findings are in line with our demand reduction model results.

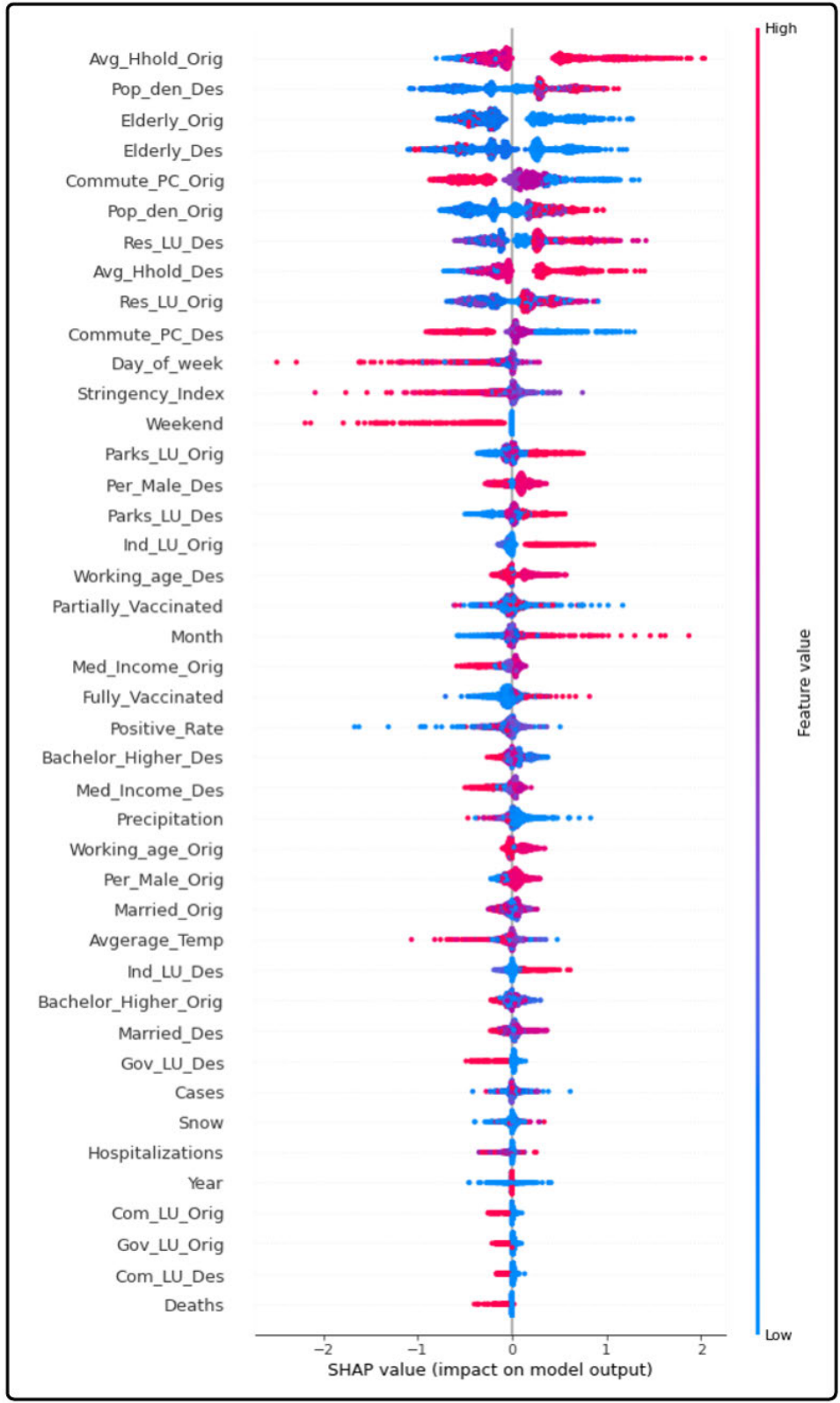

**Figure 6.** SHAP summary plot for Innisfil's direct demand model.

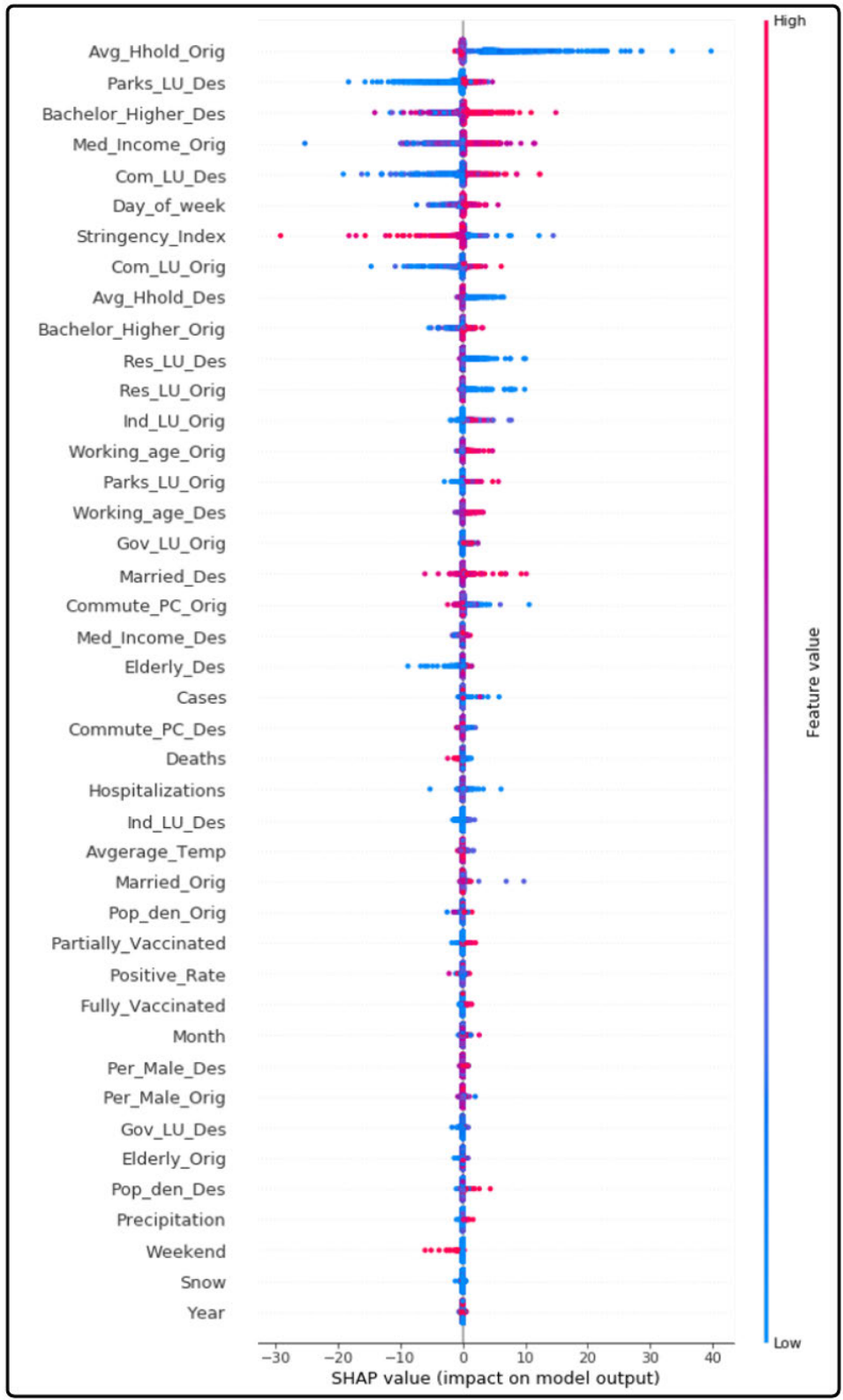

**Figure 7.** SHAP summary plot for Chicago's direct demand model.

## Discussion and concluding remarks

The on-going COVID-19 pandemic has significantly impacted individual's mobility[14,30]. Understanding these effects can help planners and policy makers to better anticipate the pandemic's future impacts, which may contribute both to recovery efforts and plan responses to additional waves of COVID-19 or other pandemics. This study analyzed and compared the impacts of the COVID-19 pandemic on the demand for ridesourcing services in the Town of Innisfil and the City of Chicago. Previous literature on the topic indicated that the pandemic resulted in a substantial reduction in the ridesourcing demand[20,23,24]. Moreover, the study conducted by Yu et al.[20] found that the COVID-19 cases and policy measures adversely affected the demand for ridesourcing services in China. Our findings support these results, especially in the North American context and add to the existing body of literature by: (a) identifying the differences in the impact of the pandemic on the ridesourcing services between small towns and large cities, (b) analyzing the effects of the stringency index, COVID-19 related variables, and vaccination rates on the reduction levels, and (c) identifying the factors affecting the direct demand in the post-pandemic period.

Overall, the pandemic has affected the usage behaviour of ridesourcing services in Innisfil and Chicago. The highest reductions in the demand occurred at night and on weekends, as a consequence of closing non-essential businesses. The physical and mental health issues associated with the pandemic resulted in fewer trips in areas with higher proportions of seniors and more trips to parks and green spaces. Moreover, the use of the pandemic-related factors contributed to interpreting the temporal variations in demand for ridesourcing and understanding their effects, which highlights the importance of incorporating them in travel behaviour studies. Results revealed that the demand was adversely affected by the stringency index and COVID-19-related variables, and positively affected by vaccination rates. These findings illustrate the importance of vaccination for the recovery of ridesourcing services. However, the extent of the effects of pandemic-related factors varied between the Town of Innisfil and the City of Chicago. Compared to Innisfil, ridesourcing demand in Chicago experienced higher reductions in demand, were more affected by the number of hospitalizations and deaths, were less impacted by vaccination rates, and had lower recovery rates. Additionally, the demand in Chicago was less affected by vaccination rates than in Innisfil, explaining the slower recovery rate observed in Chicago. These variations are due to a variety of factors including differences in the use purposes, operating policy, population density, and land area between the two locations.

This research has two main limitations: (a) Innisfil's ridesourcing dataset does not cover the early stages of the pandemic (from March to August 2020), and (b) our findings are based on ridesourcing data from Innisfil and Chicago, therefore, they may not be directly generalized to other towns or cities. Future research may explore the impact of the COVID-19 pandemic on the demand for other modes of transportation as well as the supply of ridesourcing services. Another future direction could be to investigate the factors affecting the recovery of public transit and ridesourcing services.

## Data Availability

The data used in this study are publicly available, except for Innisfil's ridesourcing and land-use data. These datasets were provided under license for the current study and, therefore, are not publicly available.

## Code Availability

All analyses were performed using python 3.9 and figures were prepared using Tableau Desktop 2021.4 as well as QGIS 3.16. The code and sample data from Chicago are available at: https://github.com/LiTrans/RidesourcingDemandModels/tree/main/Code

## Acknowledgements

We are thankful to the Town of Innisfil and Uber Technologies Inc. for providing us access to the operational data of Innisfil Transit Service that was used in this study. We also thank Nicoleta Hera for her time and effort in helping us improve the quality of this manuscript.

## Author contributions statement

**N.A.** and **B.F.** conceived and designed the study as well as developed its methodologies. **N.A.** performed the analyses, created the figures and tables, and wrote the manuscript with input from **B.F.**. All authors approved the final version of the manuscript.

## Competing interests

The authors declare no competing interests.

## Additional information

Supplementary Information for: The Impact of COVID-19 Pandemic on Ridesourcing Services Differed Between Small Towns and Large Cities

# Supplementary Information for: The Impact of COVID-19 Pandemic on Ridesourcing Services Differed Between Small Towns and Large Cities

**Nael Alsaleh[1] and Bilal Farooq[1,*]**

[1]Laboratory of Innovations in Transportation (LiTrans), Department of Civil Engineering, Ryerson University, Toronto, ON M5B 2K3, Canada
[*]bilal.farooq@ryerson.ca

## Supplementary Tables

| Hyperparameters | Model | | | |
| --- | --- | --- | --- | --- |
| | Percent Reduction in Daily Demand | | Direct Demand | |
| | Town of Innisfil | City of Chicago | Town of Innisfil | City of Chicago |
| Criterion | Mean Squared Error | Mean Squared Error | Mean Absolute Error | Mean Squared Error |
| Maximum Depth | 20 | 10 | 50 | None |
| Maximum Features | log2 (No. of Features) | log2 (No. of Features) | log2 (No. of Features) | No. of Features |
| Minimum Sample Leaf | 1 | (0.0028 x Sample Size) | 1 | 1 |
| Minimum Sample Split | 2 | (0.0227 x Sample Size) | 2 | 2 |
| Number of Estimators | 100 | 100 | 150 | 100 |
| Maximum Leaf Nodes | 100 | 100 | None (Unlimited) | None (Unlimited) |
| Bootstrap | True | False | True | True |

**Supplementary Table 1.** Optimal hyperparameters set for the models developed in this study.

| Performance Metric | Model | | | |
| --- | --- | --- | --- | --- |
| | Percent Reduction in Daily Demand | | Direct Demand | |
| | Town of Innisfil | City of Chicago | Town of Innisfil | City of Chicago |
| Mean Absolute Error (MAE) | 0.09 | 0.03 | 1.24 | 0.81 |
| Mean Squared Error (MSE) | 0.015 | 0.002 | 3.49 | 4.52 |
| Root Mean Squared Error (RMSE) | 0.12 | 0.042 | 1.87 | 2.13 |
| Coefficient of Determination ($R^2$) | 0.59 | 0.93 | 0.81 | 0.85 |

**Supplementary Table 2.** Model Performance. Across all performance metrics, the models in Chicago had a higher predictive accuracy than those in Innisfil. This is most likely because Chicago's datasets contained more samples.

## Supplementary Figures

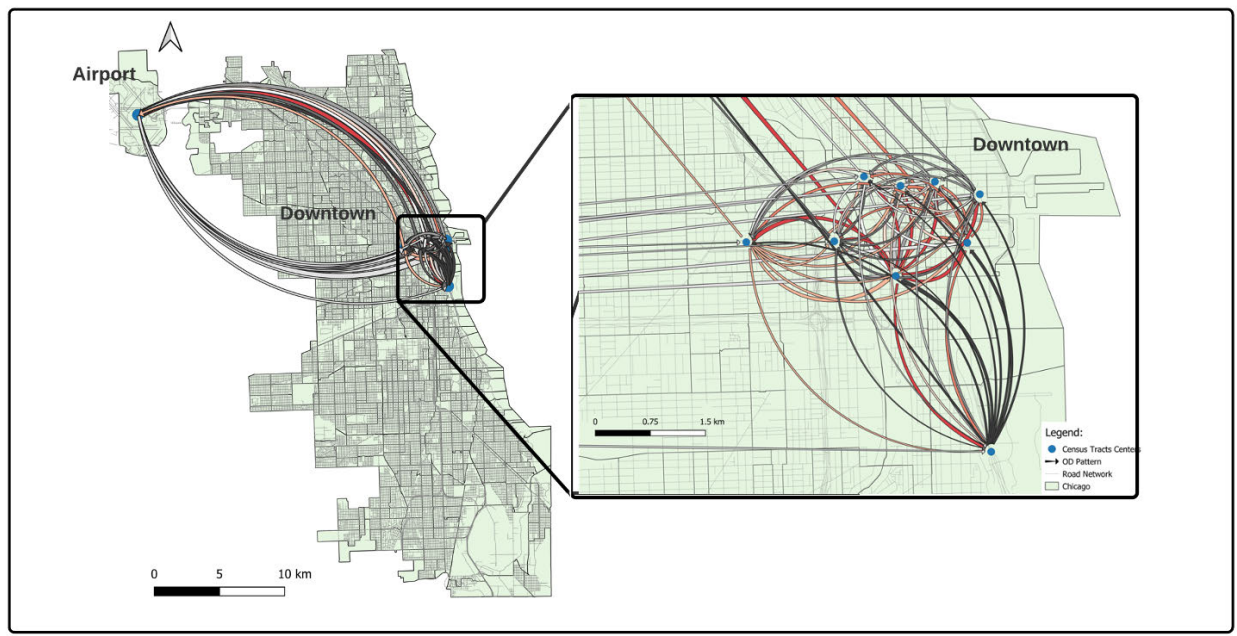

**Supplementary Figure 1.** Pre-pandemic origin-destination (OD) flow patterns for the ridesourcing services in Chicago. As during the pandemic, high-demand OD pairs were in the downtown area and between the downtown and O'Hare International Airport. The pre-pandemic trip data in Innisfil did not have the required information to develop the OD flow patterns. However, the ridesourcing trip density maps developed by Sweet et al.[1] for the time period from May 2017 to February 2020, showed that the Innisfil's central area had the highest pick-up and drop-off density, which is similar to the patterns observed during the pandemic.